\documentclass[apj]{emulateapj}
\usepackage{lscape} 
\shortauthors{Cardamone et al.}
\shorttitle{Mid-IR colors of X-ray detected AGN}
\begin{document}
\title{Mid-infrared Properties and Color Selection for X-ray Detected AGN in the MUSYC ECDF-S field}
\author{Carolin N. Cardamone\altaffilmark{1,2}, C. Megan Urry\altaffilmark{1,2}, Maaike Damen\altaffilmark{3}, Pieter van Dokkum \altaffilmark{1,2}, Ezequiel Treister \altaffilmark{4} ,  Ivo Labb$\acute{e}$\altaffilmark{5}, Shanil N. Virani \altaffilmark{1,2}, Paulina Lira\altaffilmark{6}, and Eric Gawiser\altaffilmark{7},}
\email{ccardamone@astro.yale.edu}
\altaffiltext{1}{Department of Astronomy, Yale University, New Haven, CT 06511, USA}

\altaffiltext{2}{Yale Center for Astronomy and Astrophysics, Yale University, P.O.~Box 208121, New Haven, CT 06520}
 
\altaffiltext{3}{Leiden Observatory, Universiteit Leiden, 2300 RA Leiden, Netherlands}

\altaffiltext{4}{European Southern Observatory, Santiago, Chile}
\altaffiltext{5}{Carnegie Observatories, Pasadena, CA 91101}
\altaffiltext{6}{Departamento de Astronomía, Universidad de Chile, Casilla 36-D, Santiago, Chile}
\altaffiltext{7}{Department of Physics \& Astronomy,Rutgers University, Piscataway, NJ 08854-8019}

\begin{abstract}
We present the mid-infrared colors of X-ray-detected AGN and explore mid-infrared selection criteria.
Using a statistical matching technique, the likelihood ratio, over 900 IRAC counterparts were identified with a new MUSYC X-ray source catalog that includes $\sim$ 1000 published X-ray sources in the {\it Chandra} Deep Field-South and Extended {\it Chandra} Deep Field-South.
Most X-ray-selected AGN have IRAC spectral shapes consistent with power-law slopes, $f_{\nu} \propto \nu^{\alpha}$, and display a wide range of colors, $-2 \le \alpha \le 2$.
Although X-ray sources typically fit to redder (more negative $\alpha$) power-laws than non-X-ray detected galaxies, more than 50\% do have flat or blue (galaxy-like) spectral shapes in the observed 3-8 $\mu$m band. 
Only a quarter of the X-ray selected AGN detected at 24 $\mu$m are well fit by featureless red power laws in the observed 3.6-24 $\mu$m, likely  
 the subset of our sample whose infrared spectra are dominated by emission from the central AGN region.
Most IRAC color-selection criteria fail to identify the majority of X-ray-selected AGN, finding only the more luminous AGN, the majority of which have broad emission lines.
In deep surveys, these color-selection criteria select 10-20\% of the entire galaxy population and miss many moderate luminosity AGN.

\end{abstract}


\section{INTRODUCTION}
Theoretical arguments and observational evidence indicate that Active Galactic Nuclei (AGN) may play an important role in galaxy evolution  \citep{fabian2003,fabianetal2006,crotonetal2006,krieketal2006,sijackietal2007,schawinskietal2007}.
The study of this mutual feedback requires a complete sample of AGN, their host galaxies, and comparable inactive galaxies at the same epochs.
To discover these populations, surveys with deep multi-wavelength coverage are necessary.

Today, AGN are identified at many wavelengths, each of which selects a slightly different sample.
For example, radio surveys preferentially find radio-loud AGN \citep{brinkmannetal2000}, while optical and ultraviolet-excess surveys find luminous unobscured AGN \citep{richardsetal2004,Mazzarellabalzano1986}.
Because they detect all but the most heavily obscured AGN, hard X-ray surveys
 efficiently select relatively unbiased and complete samples of AGN \citep{branthasinger2005}, even if 
X-ray background models predict a still-hidden population of Compton-thick AGN \citep{treisteretal2004,worsleyetal2005}.
For obscured AGN (roughly three quarters of the total population), the rest frame optical light is dominated by the host galaxy (e.g., \citealt{treisteretal2004,treisterurry2006}).
This obscuring gas and dust eventually re-emits the absorbed X-rays at infrared wavelengths.
In general the optical/infrared spectral shapes of AGN contain two broad components of thermal emission: the ``Big Blue Bump'' seen at optical and ultraviolet wavelengths, often attributed to emission from a hot accretion disk, visible only in unobscured AGN and an infrared excess attributed to re-radiation from circumnuclear dust on larger scales visible in all AGN in the far-infrared \citep{sandersetal1989,sanders1999,pierkrolik1992}.

Mid-infrared diagnostic studies have found strong continuum emission, originating in dust heated by the intense nuclear radiation field, an important indicator of AGN emission \citep{laurentetal2000}.
Indeed, surveys with the {\it Spitzer Space Telescope} Infrared Array Camera  (IRAC; \citealt{fazioetal2004}) are finding AGN via color-selection techniques by targeting strong red power-law continua \citep{Alonso-Herrero2006,hickoxetal2007,donleyetal2007}.
This emission can be used to find even Compton-thick AGN \citep{pollettaetal2006}.
However, many color-selection techniques developed with {\it Spitzer} data have been based on the infrared colors of optically-selected AGN samples in shallow surveys (e.g., \citealt{lacyetal2004,sternetal2005,hatziminaolgou2005}).
The ideal AGN selection technique is both highly reliable, meaning most objects selected are indeed AGN, and highly complete, meaning most AGN are identified.
To evaluate and improve infrared selection of AGN, it is important to investigate how the mid-infrared colors of X-ray-selected AGN samples compare to those of optically-selected samples.

Finding high redshift AGN in multi-wavelength surveys requires both a large area and sensitive detection limits.
The  MUltiwavelength Survey by Yale-Chile (MUSYC\footnote[1]{\url{http://www.astro.yale.edu/MUSYC}}; \citealt{gawiseretal2006}) is a square-degree survey of four fields to
limiting depths of $R_{AB}=26$ and $K_{AB}=22$ ($K_{AB}=23$ in the central areas; \citealt{quadrietal2007}), with extensive
follow-up spectroscopy. 
One MUSYC field, the Extended {\it Chandra} Deep Field-South (ECDF-S),
has been surveyed by the {\it Chandra} X-ray observatory \citep{giacconietal2002,alexanderetal2003,lehmeretal2005,viranietal2006}.
The sensitivity of the  {\it Chandra} data allows for
detection of luminous AGN out to large redshifts\footnote[2]{Luminous AGN ($L_X \sim 10^{44} \: {\rm ergs \: s^{-1}}$) can in principle be detected to very high redshifts  in the ECDF-S ($z \sim 10$), but the volume sampled is too small to make such a detection likely.}
and more modest luminosity AGN ($\sim 10^{42-44} \: {\rm ergs \: s^{-1}}$) back to the quasar era ($z\sim2$, \citealt{branthasinger2005}).
Recently, the existing MUSYC data in the ECDF-S were augmented with very deep IRAC {\it Spitzer} data in the SIMPLE\footnote[3]{\url{http://www.astro.yale.edu/dokkum/SIMPLE/}} survey (Damen et al. 2008, in prep.) and by the GOODS and SWIRE surveys at 24 $\mu$m \citep{dickinson2003,lonsdale2003}.
These two surveys are among the deepest {\it Spitzer} observations.
This results in a large sample of X-ray-selected AGN with mid-infrared coverage, unique in the combination of area and X-ray and infrared depth.

In this paper we evaluate mid-infrared selection techniques using the X-ray selected sample of AGN in the ECDF-S.
In Section 2, we describe the X-ray and infrared data, detail the likelihood ratio-matching technique used to find {\it Spitzer} counterparts for each X-ray source, and present the infrared identifications.
We investigate the mid-infrared colors of these X-ray sources in Section 3.
In Section 4 we
compare our sources to published IRAC-selection criteria and assess the impact of adding the 24 $\mu$m data to the IRAC bands.
In Section 5, we discuss why the majority of our X-ray-selected AGN would be missed by current infrared selection criteria.
Conclusions are given in Section 6.
Throughout this paper we assume $H_0 = 71  \: {\rm km \: s^{-1} \: Mpc^{-1}}$, $\Omega_{\rm m} = 0.3$ and $\Omega_{\rm \lambda} = 0.7$. 

\section{\uppercase{OBSERVATIONS \& ANALYSIS}}
\subsection{{\it Chandra} Data}
Three X-ray surveys have overlapping coverage of the same region of the sky:
the very deep central {\it Chandra} Deep Field-South (CDF-S; \citealp{giacconietal2002}, hereafter G02; \citealt{alexanderetal2003}, hereafter A03), the deep Extended {\it Chandra} Deep Field-South (\citealp{lehmeretal2005}, hereafter L05; \citealp{viranietal2006}, hereafter V06), and the shallower XMM pointing that covers this region \citep{streblyanska2004}.
For this study, we use only the two {\it Chandra} surveys because the higher resolution of {\it Chandra} allows for fainter point source detections.  
In contrast, the XMM data are better suited for X-ray spectroscopy of the brighter sources because of the larger effective area \citep{streblyanska2006,dwellypage2006}.
The CDF-S consists of 1 Ms of data reaching a limiting on-axis flux of $5.2\times10^{-17} \: {\rm ergs \: s^{-1} \: cm^{-2}}$ ($2.8\times 10^{-16} \: {\rm \: ergs \: s^{-1} \: cm^{-2}}$) in the 0.5-2.0 keV (2.0-8.0 keV) band (A03).
The 11 individual pointings of the CDF-S cover a central area of $\sim 300$ arcmin$^2$  (G02).
The four 250-ks {\it Chandra} pointings of the ECDF-S overlap the deep central region, covering a total area of $\sim 1100$ arcmin$^2$.  
On axis the ECDF-S reaches a limiting flux of $1.7\times 10^{-16} \: {\rm \:ergs \:cm^{-2}\: s^{-1}}$ ($3.9 \times 10 ^{-16} \: {\rm \:ergs \: s^{-1} \:cm^{-2}}$) in the 0.5-2.0 (2.0-8.0) keV band (V06).
We combine published point source detections from these fields into a new unified MUSYC X-ray point source catalog.

We begin with the most conservative detection catalogs for the CDF-S and ECDF-S, A03 and V06. 
Since the CDF-S goes significantly deeper in flux than the ECDF-S and covers a much smaller area between the four ECDF-s pointings, we expect that only a fraction of sources will be common to both surveys.
Both catalogs used similar detection methodologies: first filtering the data to eliminate periods of high X-ray background, then using {\it wavdetect} \citep{freemanetal2002} for source selection,  which is robust in detecting individual sources in crowded fields.
The MUSYC X-ray catalog first adopts the 326 sources presented in A03 Table 3.
We next search for new sources in V06 Tables 4 and 5 as described in what follows.
Table 4, the primary catalog in V06 uses the same conservative false-positive probability threshold for detection ($p_{\rm thresh}=10^{-7} $) as \citet{alexanderetal2003}.
Given this detection threshold, we expect no more than 2 or 3 false X-ray sources in either A03 or Table 4 of V06.
Table 5 of V06, with $p_{\rm thresh}=10^{-6} $, allows for a larger percentage of false positives but includes a greater number of bona-fide X-ray sources in the catalog that would otherwise be missed.
In order to improve the absolute astrometric solutions in the C06 catalogs, small shifts were applied to the X-ray sources RA and DEC, $\le 1$ arcsecond.
The shifts were calculated for each of the four pointings in the ECDF-S using optical counterparts in the MUSYC BVR image.
No shift was necessary to align A03 catalog positions.
For each source in V06, we calculated the distance to the nearest X-ray sources in A03 and defined combined X-ray positional uncertainties for each pair by  $\sigma=\sqrt{\sigma_{A03}^2+\sigma_{V06}^2}$, where $\sigma_{XR}$ is the 1 $\sigma$ error of a Gaussian based on the positional errors reported in each catalog.
We included in our MUSYC X-ray catalog the 503 sources from V06 that were more than $4\sigma$ ($\sim 3$ arcseconds) from any source in A03.
The choice of $4\sigma$ is conservative and was determined by detailed inspections of both the X-ray and MUSYC BVR images.
The median separation between the reported source positions for the 148 X-ray sources found in both A03 and V06 was 0.58 arcseconds.
A total of 829 unique X-ray sources resulted from the union of A03 and V06, the first 771 of which were detected with the conservative false-positive probability threshold.

We next investigated the other published X-ray catalogs in this field for the same two data sets. 
\citet{lehmeretal2005}, created a larger source catalog for the ECDF-S region by retaining data obtained during small X-ray background flares and using the same lower significance threshold parameter as Table 5 in V06.
A very small $\le 1$ arcsecond positional shift was applied to the L05 catalog to align the astrometry to the MUSYC BVR image.
Using the X-ray positional uncertainties, we found 141 additional X-ray sources in L05.
Finally, \citet{giacconietal2002} used a set of unique detection methods including a modified version of Sextractor \citep{bertinarnouts1996} and running {\it wavdetect} with different selection criteria.
Three catalogs are presented in G02: Table 2 containing sources common to both detection methods, Table 3 containing Sextractor sources only and Table 4 containing {\it wavdetect} sources only.
A slightly larger fixed positional shift ($\sim 1$ arcsecond) was necessary to align the X-ray sources presented in G02 with our MUSYC BVR image.
Using X-ray positional uncertainties, we found a total of 47 additional X-ray sources not previously presented in A03, V06 or L05: 17 from Table 2, 19 from Table 3 and 10 from Table 4.
Although there is a good possibility that some of these additional X-ray sources may be spurious, those will fail to match to an infrared counterpart and thus will not affect the conclusions of this paper (see \S~2.3).

The complete MUSYC X-ray catalog contains 1017 individual sources whose X-ray properties are given in columns 1 through 5 in Table 1.  
Column 1 presents a new unique MUSYC X-ray ID number for each source.  
Column 2 presents the original X-ray catalog and catalog ID for each X-ray source (e.g., V06:323 is source number 323 from Virani et al. 2006).  All X-ray source properties, such as flux or count rate, are taken from the referenced detection catalog.
Columns 3 and 4 present the RA and DEC, J2000, for each X-ray source, based on the MUSYC BVR image.  
They are on a common system incorporating the small $\Delta$RA and $\Delta$DEC applied to each catalog to shift the coordinates to agree with the MUSYC BVR image.
Column 5 gives the hardness ratio, defined by $\frac{H-S}{H+S}$, where H and S are the X-ray counts in the hard (2-8 keV) and soft (0.5-2 keV) bands, respectively.  For sources detected only in the hard band (soft band), the hardness ratio is set equal to 1 (-1).  For sources detected only in the full band (0.5-8 keV), the hardness ratio is set equal to 10.  G02 defined the hard band as 2-10 keV.  Therefore, we adjusted the published hard band counts from G02 to equal those that would have been detected at 2-8 keV, the hard band defined by the other three authors, using PIMMS (version 3.9c; \citealt{mukai1993}) with the average X-ray spectrum measured by G02 ($\Gamma=1.4$; $H_{2-8}=0.9896 \times H_{2-10}$).

\subsection{{\it Spitzer} Data}
Considerable {\it Spitzer} time has been invested in the ECDF-S with both IRAC and MIPS (the Multi-Band Imaging Photometer for {\it Spitzer}; \citealt{riekeetal2004}).
The GOODS Survey along with additional Guaranteed Time Observer (GTO) time gathered deep IRAC and MIPS data on the central region of this field \citep{dickinson2003}. 
Additionally, the {\it Spitzer} Wide-area InfraRed Extragalactic survey (SWIRE; \citealt{lonsdale2003}) covers a nearly 8 square degree region around the original {\it Chandra} Deep Field-South in a shallow survey mode with both the MIPS and IRAC cameras.  
Finally, a public deep IRAC survey was conducted in the region covering the ECDF-S (SIMPLE; PI. P. van Dokkum).

The IRAC data cover $3-8$ $\mu$m to $5 \: \sigma$ limiting depths of 24.4 (3.6 $\mu$m), 24.0 (4.5 $\mu$m), 22.2 (5.8 $\mu$m) and 22.2 (8.0 $\mu$m) in AB magnitude.
In units of flux density, these values are 0.63 $\mu$Jy, 0.91 $\mu$Jy, 4.7 $\mu$Jy, and 4.7 $\mu$Jy, respectively.
A catalog of more than 60,000 sources was created with SExtractor using a weighted sum of the 3.6 and 4.5 $\mu$m data (Damen et al. 2008, in prep.).
Of these, we have reliable photometry in all four IRAC bands for $38,755$ sources (3$\sigma$), which will be used in this study.
The astrometry was calibrated using the MUSYC BVR detection image ($R_{AB}=27.1$);
the resulting positional accuracy for individual sources is $\lesssim 0.3$ arcseconds ($1 \sigma$; Damen et al. 2008, in prep.).

We also use the SWIRE 24 $\mu$m catalog provided in data release 3 \citep{lonsdale2003,surace2005}, which  cover the entire area of the X-ray data to AB magnitude 18 (0.23 mJy), and the GOODS data in the central region, which goes to $\sim5$ magnitudes deeper \citep{dickinson2003}.
We adopt the aperature fluxes from the SWIRE catalog which agree with the published GOODS fluxes to better than 5\%.
The positional accuracy of the 24 micron data is considerably less than that of the IRAC data;
however, all 24 micron source positions were matched to IRAC sources, decreasing the positional uncertainties.

 \subsection{Catalog Matching}
Following the work of \citet{brusa2005,brusaetal2007}, we use the Likelihood Ratio (LR), a statistical matching technique, 
to find {\it Spitzer}  counterparts to the ECDF-S X-ray sources.
Although the median positional accuracy of {\it Chandra} is $\sim 0.\arcsec6$, off axis the PSF broadens and becomes circularly asymmetric, 
leading to larger positional uncertainties in faint off-axis sources.
The positional accuracy for each detected {\it Chandra} source depends on the number of counts detected and the distance of the source from the aim point.
For several X-ray sources, this positional uncertainty encompasses multiple potential infrared counterparts, while for others there may be no counterpart to the survey detection limits.
Due to the high source density of the faintest IRAC sources, there is a non-negligible random chance of
finding a faint IRAC object near an unassociated  X-ray source.
Simulations showed that $\sim$ 60\% of random positions in the field fall within 4$\arcsec$ of an IRAC source, the maximum distance for which we find an IRAC counterpart for an X-ray source.
The Likelihood Ratio takes into account the background object density and the positional accuracy of both the {\it Spitzer} and the {\it Chandra} data, and returns the probability that a given match is the correct one.
A nearest neighbor approach, restricted to a typical search radius of 1.5 arcseconds, would miss 109 IRAC counterparts that we recover with our technique.
Although larger search radii can return more possible counterparts, a nearest neighbor approach does not indicate which counterparts fall inside the search area by chance and it can only distinguish among multiple possible candidates by their distance from the reported X-ray source position. 
Our method weights the distance using the positional errors and adds a prior on the brightness of the prospective counterpart.

The Likelihood Ratio is the ratio between the probability of finding the true infrared
counterpart at a given distance from a particular X-ray source divided by the probability of finding a
background object at that distance (see \citealt{sutherlandsaunders1992,ciliegietal2003}).
When matching the {\it i}th X-ray source, to a possible {\it j}th infrared counterpart, we calculate:
\vspace{-0.2cm}
{\small{
\begin{equation} 
LR_{ij}=\frac{q(m_j) f(r_{ij})}{n(m_j)},
\end{equation}}}

\noindent where {\it q(m)} is our empirically determined probability 
distribution for infrared counterparts as a function of magnitude; 
{\it f(r)} is the probability distribution function 
of the positional errors in source {\it i} and counterpart {\it j} (modeled as a two dimensional Gaussian defined with $\sigma_{gaussian}=\sqrt{\sigma_{X,i}^2+\sigma_{IR,j}^2}$);
and {\it n(m)} is the surface density of background objects as a function of magnitude.  
When calculating the probability distribution function 
of the positional errors, $\sigma_{X,i}$ is unique to each X-ray source depending on its off axis angle and brightness, while no such field-dependent distinction is made for the infrared counterparts.
We adopt $\sigma_{IRAC} = 0.3 \arcsec$ and $\sigma_{24\mu m} = 0.5 \arcsec$ (\S~2.2).
Both {\it q(m)} and {\it n(m)} depend on the depth of the infrared data. 
This technique is inherently Bayesian, and the resulting likelihood ratio 
assumes the priors {\it q(m)}, {\it n(m)} and {\it f(r)}  are known.  

When calculating the background counts of infrared sources, {\it n(m)}, 
we correct for the fact that there are actually fewer faint sources near X-ray objects than near random positions in the field \citep{brusaetal2007}.
This follows from the high probability of finding a bright infrared counterpart near any given X-ray position 
and the fact that the fainter sources very close to these bright counterparts go undetected.
To do this we first estimate the background around a sample of IRAC objects with a magnitude 
distribution similar to those IRAC objects located within $1 \arcsec$ of the X-ray source positions;
the positional uncertainties make it highly probable that IRAC objects within this distance are true counterparts.
Subtracting this estimated background magnitude distribution from the magnitude distribution of the actual IRAC sources near the X-ray source positions
provides a better estimate of the magnitude distribution of IRAC sources associated with X-ray sources.
Experimentation showed that using this improved estimate was stable to multiple iterations.
We then estimate the background around a sample of IRAC sources with this final magnitude distribution for objects associated with X-ray source positions; this is the {\it n(m)} used in our likelihood ratio calculations. 

Next, we count the total number of infrared objects lying within a distance $r$ of all X--ray sources, $N_{total}(m)$.
Then we multiply the background distribution by the area searched to find these sources; the latter is  the number of X--ray sources, $N_{X}$, times the area of each circle searched, $\pi  r^2$.
The difference between $N_{total}(m)$ and the expected background, $n(m) N_{X}  \pi  r^2$, is our expected distribution of real counterparts, $N_{real}(m)$.  
It is important to understand this is a statistical quantity; rather than predicting the expected infrared counterpart for any individual X--ray source, it calculates the expected magnitude distribution for all $N_{X}$ sources.  
With our expected distribution of real counterparts we derive the normalization, $Q$, as the ratio of the total expected number of real counterparts, $\sum{N_{real}(m)}$, divided by the total number of X--ray sources, $N_{X}$.
Then we compute our expected distribution of true infrared counterparts as a function of magnitude:
\begin{equation}
q(m)=\frac{N_{real}(m)}{{\displaystyle{\sum_{m}}}{N_{real}(m)}} Q 
\end{equation}
Finally, we compute the likelihood ratio for each infrared object located near an X--ray source using Equation 1.  

When we search for infrared counterparts over a larger radius we see more possible counterparts, but the large distance between these prospective counterparts and the X--ray source drives down the likelihood ratio.  
We must then carefully select the best likelihood ratio which can distinguish between true matches and false matches.
Because the presence of multiple prospective counterparts provides additional knowledge to the likelihood ratio, we use the reliability ($R$) to select counterparts \citep{sutherlandsaunders1992}.
The reliability is the probability that a given infrared counterpart is the correct infrared counterpart for each X-ray source.  
For each possible infrared counterpart {\it j}, we calculate the reliability by:
\begin{equation}
 R=\frac{LR_j}{\sum{LR_i} + (1-Q)}
\end{equation}
Here, ${\sum{LR_i}}$ is the sum over all possible infrared counterparts for a given X-ray source.
The $(1-Q)$ term is the probability of there being no infrared counterpart.
The calculated reliabilities form a bimodal distribution, clustering at values above R=0.9 and below R=0.1, and dividing cleanly at R=0.6.
Therefore, we accept all matches with $R \ge 0.6$.
The likelihood ratio approach assumes exact knowledge of the priors,  {\it q(m)} and {\it n(m)}.  
Because we estimate these values from the data, the likelihood ratio does not reflect uncertainties in our estimation of the background.
However, with a threshold value of R = 0.6, we find from Monte Carlo simulations of the background estimation that fewer than 1\% of our matches are affected by this uncertainty.

\subsection{{\it Spitzer}  Counterparts}
We find that more X-ray sources in the ECDF-S are detected with {\it Spitzer} than at any other wavelength.
For the full MUSYC X-ray catalog of 1017 sources, we find 921 {\it Spitzer} counterparts in the IRAC data: a 90\% recovery rate.
Of the 921 sources, 908 have $3\sigma$ detections in all four of the IRAC channels. 
For many of the X-ray sources, blending and confusion of IRAC sources, sometimes due to proximity to a very bright source, are responsible for the lack of IRAC counterparts.
Excluding regions where confusion or bright stars interfere, we recover infrared counterparts for 98\% of the high-confidence X-ray source detections, i.e. the subset of the X-ray sources detected in the ECDF-S with the conservative false-probability threshold.
For the remaining 2\% of X-ray sources, either an IRAC source is seen on the image near the X-ray position, but this IRAC source is not above the source detection limit of the IRAC catalog (2 cases), or between 2 and 5 IRAC sources are located at larger distances from the X-ray position (2-4 arcseconds) and no single source has a large enough probability of being the correct counterpart (14 cases).
We find no convincing case of an undetected high significance X-ray source in the IRAC image.
That is, essentially all X-ray sources in our catalog are bright enough in the infrared to be detected in the deep IRAC data.

There are 453 
unique MIPS counterparts combing the results from the 24 micron surveys in this field.
We find 181 X-ray counterparts in the SWIRE catalog and 318 counterparts in the GOODS catalog from the central region.  
Combining the counterparts found in the MIPS and  IRAC data, we have 428 sources with complete SEDs from 3.0 - 24 microns.

All matched counterparts and their mid-infrared fluxes are presented in Table 1.  
Columns 7, 8, 9 \& 10 give the observed flux in the 3.6, 4.5, 5.8 \& 8.0 $\mu$m IRAC bands, respectively. 
The flux units are in $\mu $Jy, and are followed by the 1$\sigma$ errors in parentheses.
Column 11 gives the MIPS 24 $\mu$m flux in $\mu$Jy followed by the 1$\sigma$ error in parentheses. The number is preceded by a G if it comes from the deeper GOODS data and an S if it is taken from the shallower SWIRE survey.

\subsection{Redshifts}
Although the majority of X-ray sources turn out to be AGN, it is also possible for deep X-ray surveys to detect Galactic stars as well as galaxies that are vigorously forming stars.
Luminosities above $10^{42} \: {\rm \:ergs \: s^{-1}}$ separate out sources that are primarily powered by black hole accretion rather than star formation \citep{persic2004,liraetal2002}.
In order to study intrinsic X-ray luminosities
and to apply a luminosity cut to exclude Galactic stars and starburst galaxies from our AGN sample,
we require some knowledge of a source's distance or redshift.
To this end, we are obtaining optical spectra of our X-ray detected sources using Magellan IMACS and VLT/VIMOS (Treister et al. 2008, in prep).
Thus far, we have 281 spectra for our X-ray sample, 110 with secure redshifts and 113 with spectral type identifications: 
42 broad-line sources (BLAGN), 47 non-broad line luminous ($\log L_{X} \ge 42 \: {\rm \:ergs \: s^{-1}}$) X-ray sources (non-BLAGN),
15 galaxies ($\log L_{X} \le 42 \: {\rm \:ergs \: s^{-1}}$) showing no AGN feature in their spectra,
and nine stars (Treister et al. 2008, in prep.).
The BLAGN are typically referred to as Type 1 AGN, while sources showing narrow-lines are referred to as Type 2 AGN.
However, these spectral signatures are often hidden in obscured AGN and moderate luminosity AGN with large host galaxies \citep{moranetal2002,severgninietal2003}.
Therefore, we define non-BLAGN to be all sources showing narrow emission lines suggestive of Type 2 AGN emission plus those sources with $\log L_{X} \ge 42 \: {\rm \:ergs \: s^{-1}}$ that show galaxy-like optical spectra.
To the 110 redshifts we obtained with IMACS, we add an additional 173 sources with spectroscopic redshifts published in the literature \citep{szokolyetal2004,croometal2001,lefevreetal2004,vanzella2005}.
We further include spectroscopic identifications from \citet{szokolyetal2004}, which adds 33 BLAGN and 74 non-BLAGN.
Another 274 sources have high quality photometric redshifts from the COMBO-17 Survey \citep{wolfetal2004}.
We note that our sample of AGN spans a range of $L_X(0.5-8 \: {\rm keV}) \sim 10^{42-45}\: {\rm \:ergs \: s^{-1}}$, typical for deep X-ray surveys, different from the shallower survey sources which are on average more luminous.
This makes a total of 557 X-ray sources (60\% of the counterpart sample) for which we can calculate the X-ray luminosity in the observed 0.5-8 keV band:
of these 416, 76\%, have $L_X(0.5-8 \: {\rm keV}) \ge 10^{42} \: {\rm \:ergs \: s^{-1}}$ and thus are bona-fide AGN.

In Table 1, we present the X-ray luminosities and spectral identifications.
Column 12 gives the observed 0.5-8 keV luminosity calculated from the redshift. For sources from G02, no 0.5-8 keV flux was reported.  Therefore, we adjusted the published 2-10 keV counts from G02 to equal those that would have been detected at 0.5-8 keV, using PIMMS (version 3.9c; \citealt{mukai1993}) with the average X-ray spectrum measured by G02 ($\Gamma=1.4$; $F_{0.5-8}=1.13 \times F_{2-10}$).
Column 13 gives the source of the redshift estimate used to calculate the X-ray luminosity as follows: 0 for VLT/FORS2 Spectroscopy Version 1.0, 1 for VIMOS VLT Deep Survey (VVDS) Version 1.0, 2 for \citet{szokolyetal2004}, 3 for \citet{croometal2001}, 4 for redshifts determined from the MUSYC spectroscopy (Treister et al. 2008, in prep.), 5 for COMBO-17 \citep{wolfetal2004}.  Those counterparts falling into the categories of BLAGN and non-BLAGN are also indicated.

\section{\uppercase{The Mid-Infrared Colors of X-ray Sources}}
\begin{figure}
\plotone{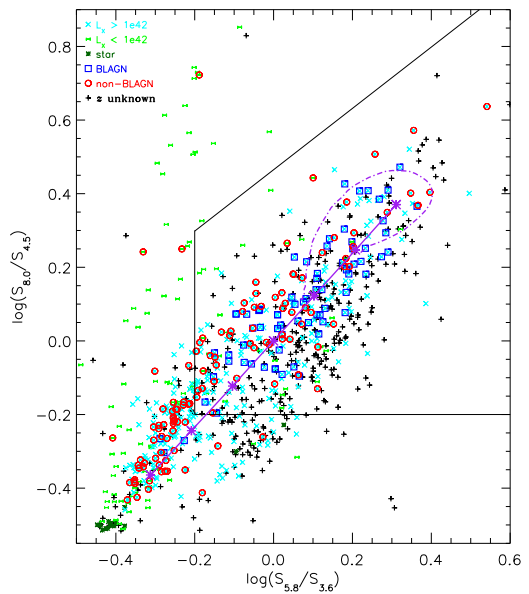}
\caption{Color-color plot for the MUSYC X-ray sources.
Also shown are two SED tracks: 
(1) the mean type 1 quasar SED from Richards et al. (2006)  red-shifted from z=0 to z=6 ({\it purple dot-dashed line, CCW});
(2) pure power-law SEDs, $f_{\nu} \propto \nu^{\alpha}$ ({\it thick purple line}), with stars marking power-law exponents of $\alpha = +1.5$ to $\alpha = -1.5$ ({\it bottom left to top right, $\Delta \alpha=0.5$}).
X-ray-selected AGN ({\it light blue crosses}) have colors consistent with power-law SEDs from the bluest to reddest colors shown here.
The black lines indicate the region suggested by Lacy et al. (2004) for selecting AGN, which excludes 42\% of X-ray-selected AGN ($L_X \ge 10^{42} \: {\rm \:ergs \: s^{-1}}$).}
\end{figure}

From 3.6 to 8.0 $\mu$m X-ray detected sources show a wide range of colors.
In Figure 1, a color-color plot of the infrared counterparts to the MUSYC X-ray catalog, sources with blue colors are in the lower left region and those with red colors are located in the upper right.
A linear relation extends from blue to red along a track occupied by power-law SED shapes ($f_{\nu}\sim \nu^{\alpha}$).
Along this track, stars indicate $\alpha$ values of +1.5 (lower left) to -1.5 (upper right) in steps of 0.5. 
For each source, X-ray luminosities are indicated where known: 
$L_X \ge 10^{42} \:{\rm \: ergs \: s^{-1}}$ plotted with light blue $\times$ symbols, 
$L_X \le 10^{42} \:{\rm \: ergs \: s^{-1}}$ plotted with light green dashes,
and those with unknown redshifts plotted with black + signs.
It is clear that the X-ray-selected AGN, $L_X \ge 10^{42} \:{\rm \: ergs \: s^{-1}}$, extend along the full range of power-law shapes.
We also explore the infrared colors of the X-ray counterparts identified as BLAGN (dark blue boxes) and non-BLAGN (red circles).
For comparison, we plot the mean Type 1 quasar SED compiled by Richards et al. (2006), from redshift 0 to 6 (dashed purple line, CCW).
Unsurprisingly the BLAGN lie near the Type 1 quasar track and have the reddest infrared colors.
Still, many AGN have significantly bluer colors than the Type 1 track, falling down and to the left on Figure 1.
In contrast, the sources classified as non-BLAGN are roughly evenly split between red ($\alpha \le 0$) and blue ($\alpha \ge 0$) colors.
Overall, while the optically identified BLAGN tend to have the reddest infrared colors, the X-ray-selected AGN population is {\it not all} red in color.

Next, we compare the IRAC colors of X-ray sources (908) to the $\sim$40,000 IRAC sources showing no X-ray emission to the limits of the published X-ray surveys, most of which are inactive galaxies. 
In Figure 2 we display two popular IRAC color-color plots; the boxes in the right and left panels correspond to two different proposed selection techniques (\S~4) and the contours enclose 75\%, 50\%, 25\% and 15\% of the IRAC sources.  
The left panel shows $\log{({S_{5.8}})/({S_{3.6}})}$ versus  $\log{({S_{8}})/({S_{4.5})}}$, where S is the observed flux in $\mu$Jy, and on the right we plot Vega magnitudes: $[5.8]-[8.0]$ vs $[3.6]-[4.5]$.
For clarity we plot only $L_X \ge 10^{42} \:{\rm \: ergs \: s^{-1}}$ sources and indicate only the BLAGN. 
On both plots, the power-law SED track and Type 1 QSO SED track are repeated from Figure 1.
Although a larger percentage of the AGN have red colors than the galaxies, in neither plot are the X-ray selected AGN clearly separated from the rest of the galaxies.

\begin{figure}
\begin{center}
\plotone{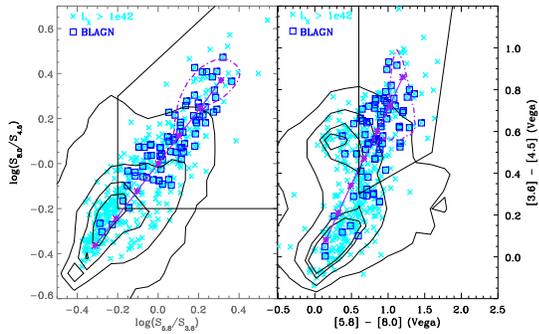}
\end{center}
\caption{
Color-color plots for all sources in the MUSYC catalog (contours indicate 75\%, 50\%, 25\% and 15\% of the IRAC sources).
The non-X-ray detected galaxies and X-ray-selected AGN are not well separated in this color space. 
{\it left}
The region defined by \citet{lacyetal2004} includes nearly 20\% of all IRAC-detected galaxies.
Therefore, this selection region is not effective at identifying a sample of AGN.
{\it right}
The selection box of \citet{sternetal2005} uses a different color-color plot of the IRAC bands.
 Only $\sim 40$\% of X-ray-selected AGN ($L_X \ge 10^{42}\: {\rm \:ergs \: s^{-1}}$)  fall in this region.
Because it eliminates 90\% of the general population of MUSYC sources, this selection box is more effective at indicating potential AGN, but misses a larger fraction of the X-ray selected AGN.
X-ray selected AGN occupy a far larger region in mid-infrared color-space than optically-selected BLAGN.}
\end{figure}

Because the galaxies and AGN appear to lie along the locus of power-laws, we fit the four observed IRAC fluxes, 3.6-8.0 $\mu$m, to power-law spectral energy distributions. 
Within the short baseline of the IRAC camera, nearly all of our sources are consistent with power laws, especially given the photometric uncertainties.
Each source with detections in all 4 IRAC channels was fit by $f_{\nu} \propto \nu^{\alpha}$; 94\%
of all sources and 97 \% of the X-ray-selected AGN had acceptable fits by power laws ($\chi^2$, $p\ge0.05$).
In Figure 3, we compare the histogram of the power-law indices, $\alpha$, for the X-ray-selected AGN and other sources.
Both types of sources show a very broad range in $\alpha$, but the AGN tend to be slightly redder than normal galaxies (more negative $\alpha$).
A KS test rejects the hypothesis that the two distributions come from the same population (P=0.0014).
While only 17\% of the galaxy sample have $\alpha \le 0$, 37\%
of  $L_X \ge 10^{42} \: {\rm \:ergs \: s^{-1}}$ sources fit a negative power-law.
Optically-selected AGN spectra show red continuum emission at infrared wavelengths, with $\alpha \sim -1$ (e.g., \citealt{elvisetal1994}).
This is consistent with our finding that 100\% of the BLAGN have acceptable power-law fits, 72\%  with $\alpha \le 0$ (Dark blue cross-hatched histogram).
As a whole, our power-law fits to the BLAGN are consistent with featureless negative power-law slopes in the mid-infrared.
However, we find a broad range of $\alpha$ values even for the BLAGN.
Similarly, the non-X-ray detected galaxies are bluer as a whole than the AGN population, even considering that the sources with the reddest $\alpha$ in the X-ray-selected AGN histogram may be broad line objects.
Indeed, although the X-ray-selected AGN have on average slightly more negative $\alpha$ values than galaxies, over half (59\%) of X-ray-selected AGN are fit by power laws with positive exponents similar to those of the bulk of the galaxy population.
Therefore, we find that most X-ray-selected AGN do not show red continuum emission at observed {\it Spitzer} IRAC wavelengths.

\begin{figure}
\begin{center}
\plotone{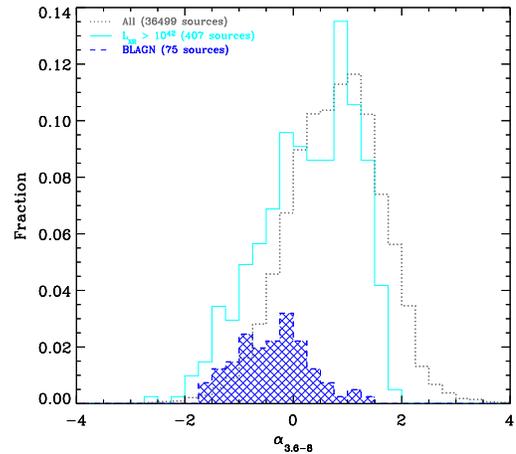}
\end{center}
\caption{
Histograms of power-law index, $\alpha$ ($f_{\nu} \propto \nu^{\alpha} $), for fits to the {\it Spitzer} IRAC photometry.
The X-ray sources with $L_{X} \ge 10^{42} \: {\rm \:ergs \: s^{-1}}$ ({\it solid light blue line}) and the subset of X-ray sources spectroscopically identified as BLAGN ({\it dark blue cross-hatched histogram}) have been normalized to the total number of X-ray sources with good estimates of redshift and good power-law fits (396). 
Also plotted is the normalized histogram for all  MUSYC sources well fit by power laws ({\it dotted black line}, 38988).
In total, 94\% of all galaxies and 97\%  of the $L_X \ge 10^{42} \: {\rm \:ergs \: s^{-1}}$ sources are consistent with a power-law SED.
A KS-test rejects the hypothesis that these two distributions come from the same population (P=0.0014).
Although the AGN are slightly redder than the galaxies on average, both the AGN and the galaxies are fit by a broad range of power-law slopes.}
\end{figure}

\section{\uppercase{Selecting AGN by Infrared Colors}}
Although selecting AGN through X-ray emission is relatively unbiased, 
current X-ray surveys may still be missing AGN that are heavily obscured by large columns of dust and gas.
All AGN, whether obscured or unobscured at optical wavelengths, reprocess and re-emit the absorbed radiation in the mid- and far-infrared \citep{granatodanese1994,nenkovaetal2002}.
Because AGN are expected to be luminous at infrared wavelengths, several authors have developed infrared selection techniques for AGN (e.g., \citealt{lacyetal2004,sternetal2005}).
Here, we examine what fraction of X-ray-selected AGN would be selected using these techniques.

\citet{lacyetal2004} found that a sample of Sloan Digital Sky Survey quasars occupied a distinct region of mid-infrared color-color space (Figures 1, 2). 
Because the selection box was defined using luminous Type 1 quasars from the Sloan survey, we look at the subset of the X-ray sources that have been identified as broad-line AGN.
Indeed, 96\%  of these sources fall inside the selected Lacy region.
Furthermore, all 61 X-ray sources identified photometrically as quasars in the Combo-17 survey also fall inside this box.
However, only 58\%  of all X-ray-selected AGN fall inside this region.
Thus, the percentage of X-ray-selected AGN that lie inside this color selection box depends on the number of luminous Type 1 objects in the sample.
That is, this infrared color selection finds the luminous unobscured quasars, but misses a high fraction of the lower luminosity AGN, many of which look like optically normal galaxies.
The brightest AGN in our sample, $L_X \ge 10^{44.5} \: {\rm \:ergs \: s^{-1}}$, are also all located inside the Lacy selection box. 
Additionally,  18\% of all sources ($\sim 11,000$ galaxies) lie within the Lacy selection box (Figure 2).
This is 20 times the total number of X-ray sources in this region (529).
A reliable AGN selection criteria should yield a sample of sources containing mostly active galaxies.  Because so many IRAC-detected sources fall inside this selection box, it is not a reliable method for selecting AGN in deep surveys.

\citet{sternetal2005} define a similar selection region using a deeper sample of spectroscopically confirmed AGN from the Bootes field. 
In the right panel in Figure 2, we plot X-ray-selected sources on the \citet{sternetal2005} color-color plot. 
Again, we find that the majority (76\%) of our spectroscopically identified broad-line AGN fall inside this selection box, but only 40\% of X-ray-selected AGN do.
Over 90\% of the brightest AGN, $L_X \ge 10^{44.5} \: {\rm \:ergs \: s^{-1}}$, are also all located inside the Stern selection box. 
In this case fewer galaxies lie inside this box (5163, $\sim$ 8\%), but they still out number the 340 X-ray sources by a factor of $\sim 15$.
The \citet{sternetal2005} selection criteria is a slightly more reliable way to select for candidate AGN, but at the same time misses a larger percentage of the total AGN population.

Although color selection techniques may work well for shallow surveys, where a larger percentage of the AGN are bright and show broad lines, they are not as reliable in deep surveys.
These selection boxes correctly highlight the region containing luminous broad-line quasars but miss most of the AGN population,
in particular non-BLAGN and low-luminosity AGN.
It is possible that some of the galaxies in these selection regions hide luminous AGN.
However, looking at Figure 2, there does not appear to be a clear region in color-color space that selects a complete sample of X-ray detected AGN and avoids returning a large portion of the entire galaxy population.
In an effort to find a more reliable selection technique, we now add the 24 $\mu$m band data to our analysis of infrared colors.

\subsection{24 $\mu$m Data}
Adding the 24 $\mu$m data to a given source's SED more than doubles the baseline over which we can determine the AGN spectral shape.
For $\sim 25$\% of our X-ray sources, consistent power-law fits are determined from the 3.6-8 $\mu$m and 3.6-24 $\mu$m data (examples shown in Figure 4, bottom panels).
However, for many sources the best-fit power law to the IRAC data alone is much bluer than one that includes the 24 $\mu$m data (examples shown in Figure 4, top panels).
We found a variety of spectroscopically identified BLAGN and non-BLAGN with 24 $\mu$m intensities that confirm the power-law slope from the IRAC data and also those whose 24 $\mu$m fluxes imply a distinct slope.
In Figure 4, the sources on the left are identified as non-BLAGN from optical spectra while the sources on the right are BLAGN.
When adding the 24 $\mu$m flux to the fit, the median non-BLAGN slope changes from a value of $\alpha_{3.6-8\mu m}=0.2$ to $\alpha_{3.6-24\mu m}=-1.0$, while the median BLAGN slope changes from a value of $\alpha_{3.6-8\mu m}=-0.3$ to $\alpha_{3.6-24\mu m}=-1.2$.
The non-BLAGN appear bluer on average in their 3.6-8.0 $\mu$m colors, while the BLAGN are more likely to show a similar $\alpha_{3.6-8\mu m}$ and  $\alpha_{3.6-24\mu m}$.
While nearly all X-ray sources show a negative slope in their $3.6-24 \mu m$ spectrum, only for a minority ($\sim 25\%$) of X-ray sources is it a strictly monotonic rise.

\begin{figure}
\begin{center}
\plotone{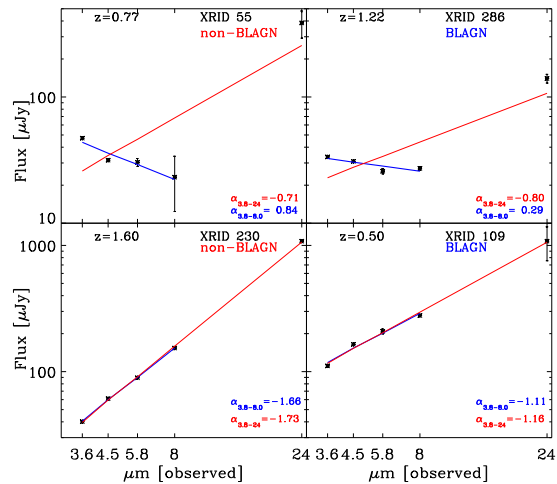}
\end{center}
\caption{
X-ray-selected AGN show a variety of spectral shapes below 8 $\mu$m, but nearly all 24 $\mu$m-detected AGN are very red between 3.6 and 24 $\mu$m.
In the top plots, a power-law fit to 3.6-8 $\mu$m photometry is significantly bluer than a power-law fit including the 24 $\mu$m flux.
In the lower plots, the same power-law fits the 3.6-8 $\mu$m and the 3.6-24 $\mu$m data.
Only $\sim 25\%$ of the MIPS sample falls into the latter category.
The plots on the left are for sources identified spectroscopically as non-BLAGN while the plots on the right are BLAGN.
Blue lines show power-law fits to $3.6-8 \mu m$ and red lines show fits including the 24 $\mu$m data point.}
\end{figure}

In Figure 5, we directly compare the fitted index $\alpha_{3.6-8\mu m}$ with that of $\alpha_{3.6-24 \mu m}$ for all 24 $\mu$m detected sources.
In this figure we include all sources whether or not their spectral shape is consistent with a power law.
The X-ray detected sources span a large range of $\alpha_{3.6-8\mu m}$ but in their longer wavelength color, $\alpha_{3.6-24 \mu m}$, they are nearly uniformly red.
In contrast, all five galactic stars detected in X-rays and at 24 $\mu$m are in the upper left corner ($\alpha_{3.6-24 \mu m} \ge 1$, dark green points).
The histogram of  $\alpha_{3.6-24 \mu m}$ peaks below -1, with $\sim$90\% of X-ray-selected AGN detected at 24 $\mu$m having $\alpha_{3.6-24 \mu m} \le 0$ (Figure 5, bottom panel).
This histogram is quite different than that for the best-fit $\alpha_{3.6-8\mu m}$ (Figure 5, right panel).
The vast majority of X-ray sources with counterparts at 24 $\mu$m have $\alpha_{3.6-24 \mu m} \le 0$: we question if this is because all AGN have negative infrared slopes, or if our shallow 24 $\mu$m coverage simply misses those AGN that would show blue colors.
We indicate sources without a 24 $\mu$m detection in the cross-hatched portion of the $\alpha_{3.6-8\mu m}$-histogram (Figure 5, right panel).
The deep GOODs data in the central portion returns 24 $\mu$m magnitudes even for sources with very blue spectral shapes in the IRAC bands; we find no significant correlation between $\alpha_{3.6-8\mu m}$ and sources lacking 24 $\mu$m detections.
Compared to  54\% of all X-ray counterparts, 48\% of the X-ray counterparts detected at 24 $\mu$m have blue spectral shapes below 8 $\mu$m ($\alpha_{3.6-8\mu m} \ge 0$).
Only 20\% of the X-ray selected AGN are consistent with a power-law SED from 3.6-24 $\mu$m ($\chi^2$, $p \ge 0.05$), but of those that fit a power law, nearly all are negative power laws (34 out of 38 sources).
This contrasts with only 43\% of the X-ray-selected AGN with IRAC power-law SEDs that have $\alpha_{3.6-8\mu m} \le 0$.
Although they have a large range in color below 8 microns, X-ray-selected AGN detected at 24 $\mu$m are uniformly red beyond 8 $\mu$m.
For comparison, we add all 24 $\mu$m detected galaxies in the GOODS survey to Figure 5 (gray points) and indicate those that are consistent with a power-law SED between 3.6 and 24 $\mu$m (purple points). 
Many of these sources may be AGN not detected at X-ray wavelengths, but the rest contain significant amounts of dust which is luminous at the observed 24 $\mu$m \citep{egamietal2004}.
The 24 $\mu$m data are not sufficient to differentiate the X-ray selected AGN from the non-X-ray detected galaxies.

\begin{figure}
\begin{center}
\epsscale{.9}
\plotone{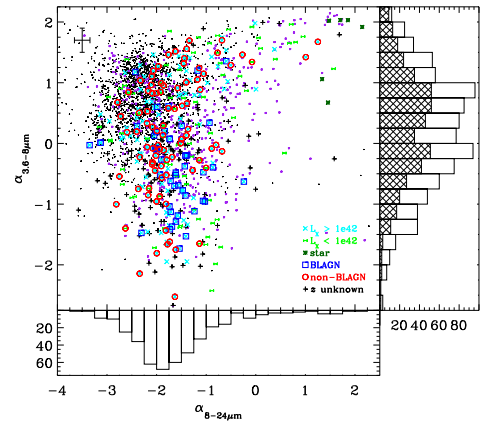}
\epsscale{1}
\end{center}
\caption{
We compare the power-law slope ($f_{\nu} \propto \nu^{\alpha} $) from the IRAC bands, $\alpha_{3.6-8\mu m}$, to the power-law slope to all 5 bands between 3.6 and 24 $\mu$m ({\it main panel}).
We include all sources whether or not $f_{\nu} \propto \nu^{\alpha} $ is a good fit to the data in order to study the approximate shape of the SED.
For comparison we plot the GOODS sample of 24 $\mu$m-detected sources ({\it gray points}) and indicate those that are consistent with a good power-law fit between 3.6 and 8 $\mu$m ({\it purple points}).
The distribution of $\alpha_{3.6-24\mu m}=$ peaks below -1.0 (bottom panel), significantly redder than either of the two peaks of the distribution of  $\alpha_{3.6-8\mu m}$ for all sources (right panel).
The cross hatched histogram indicates sources lacking 24 $\mu$m detections.
The deep GOODS data detects the 24 $\mu$m flux for sources with a large variety of IRAC spectral shapes.
The median uncertainty in $\alpha$ for the X-ray sources is indicated in the top left corner.
In the IRAC bands X-ray selected AGN show a wide range of colors, but beyond the observed 8 $\mu$m band, nearly all X-ray selected AGN fit very red spectral shapes, $\alpha \le -1.0$
However, the 24 $\mu$m flux, which highlights cold dust, does not help separate AGN from non-active galaxies. }
\end{figure}

\section{\uppercase{Discussion}}
If the AGN contribution to an X-ray source SED is a red power law at infrared wavelengths, the infrared emission from most AGN in the MUSYC sample must also include a 
significant contribution from the stars and dust in the host galaxy.
Stellar spectra are declining in the infrared, creating a blue continuum for rest wavelengths $\lambda \ge 1.6$ $\mu$m.
Because the peak of the redshift distribution for X-ray-selected AGN in the ECDF-S is at $z \le 1$, the AGN host galaxies will typically appear blue in the observed $3.6-8.0 \mu m$ band.
For the lowest redshift objects, contributions from cool dust and PAH features can also contribute to the observed mid-infrared light \citep{laurentetal2000}.
The 24 $\mu$m band is sensitive to dust emission in galaxies, making both AGN and luminous star forming galaxies easily detectable \citep{treisteretal06,egamietal2004,leflochetal2004}.

Figure 6 shows X-ray luminosity versus the fitted infrared power-law slope ($\alpha_{3.6-8 \mu m}$, bottom; $\alpha_{8-24 \mu m}$, top) for all MUSYC X-ray sources.
Although there is a trend for brighter X-ray sources to have redder power laws below 8 $\mu$m, even some of the most luminous X-ray sources in the ECDF-S ($L_X = 10^{44} - 10^{45}  \: {\rm \:ergs \: s^{-1}}$) have $\alpha_{3.6-8\mu m} \ge 0$. 
Because sources with similar $L_X$ fit such a large range of $\alpha_{3.6-8\mu m}$, it is likely that varying contributions from the  host galaxy (stars and cool dust) determine the different IRAC slopes.
This is in agreement with the large number of non-BLAGN that lie outside the infrared color selection box defined by optically-selected quasars, and with what is seen in modestly obscured AGN of Seyfert luminosities, where the host galaxy starlight can dominate the optical colors \citep{moranetal2002,cardamoneetal2007}.
In contrast, for infrared power-law selected AGN ($\alpha_{3.6-8\mu m} \le 0$), the energetics are dominated by the central AGN and thus the signatures are often visible in the optical.
Longward of 8 $\mu$m, there is no observed trend of spectral shape ($\alpha_{8-24\mu m}$) with X-ray luminosity.
Our sample agrees with other studies that find that AGN selected through IRAC power-law emission are the most luminous of those detected in X-rays \citep{donleyetal2007} and with studies that look at the most luminous hard X-ray-detected AGN and find them to have red mid-IR colors \citep{georgantopoulosetal2007}.

\begin{figure}
\begin{center}
\plotone{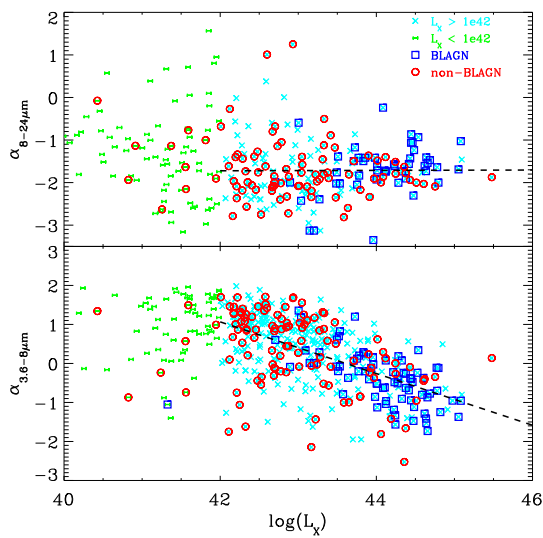}
\end{center}
\caption{
{\it Top:} At wavelengths longer than the observed 8 $\mu$m, there is no observed trend of $\alpha_{8-24\mu m}$ with X-ray luminosity.  The slope of the best fit line to the sources with  $L_X \ge 10^{42}\: {\rm \:ergs \: s^{-1}}$ is $0.04 \pm 0.06$ ({\it dashed line}).
{\it Bottom:} More luminous X-ray sources have redder infrared power laws in the IRAC bands, $\alpha_{3.6-8\mu m}$, although in all cases there is a broad range.  
The slope of the best-fit line to the sources with $L_X \ge 10^{42}\: {\rm \:ergs \: s^{-1}}$ is $-0.6 \pm 0.05$ ({\it dashed line}).
It is likely that the galaxy influence, which is proportionally larger for less luminous AGN, is the origin of this trend.}
\end{figure}

Supporting the idea that the colors below 8 $\mu$m are influenced by the host galaxy light is the fact that the X-ray spectral shape is entirely uncorrelated with the infrared spectral slope but correlated with optical AGN signatures.
We plot $\alpha_{3.6-8\mu m}$ and $\alpha_{3.6-24 \mu m}$ versus the X-ray hardness ratio, $(H-S)/(H+S)$, where H is the count rate in the hard band ($2-8$ keV) and S is the count rate in the soft band ($0.5-2$ keV) in Figure 7.
The hardness ratio is set to -1 or +1 if the source is detected in only one of the two X-ray bands, and is not shown here for detections only in the full band.
The subset of sources with optical spectroscopy is indicated by dark blue squares (BLAGN) and red circles (non-BLAGN) and sources consistent with power-law SEDs are indicated (purple stars).
There are fewer sources consistent with power-law spectral shapes between 3.6 and 24 $\mu$m, but there is no trend with X-ray hardness ratio.
This result may indicate different that the X-ray and infrared radiation is produced in vastly different environments.
The X-rays produced near the central black hole are sensitive to obscuration near the central region, while the infrared reprocessed emission at wavelengths shorter than the observed 24 $\mu$m depends on the dust distribution not only around the nucleus but throughout the galaxy.
In contrast, the optical identifications based on emission lines are somewhat correlated with hardness ratio: the left side of Figure 7, i.e., negative X-ray hardness ratios (relatively soft spectra), is preferentially occupied by BLAGN.
This reinforces the idea that the observed infrared emission is insensitive to the geometry,
or more likely, includes a substantial contribution from the host galaxy itself.

\begin{figure}
\vspace*{-10mm}
\begin{center}
\plotone{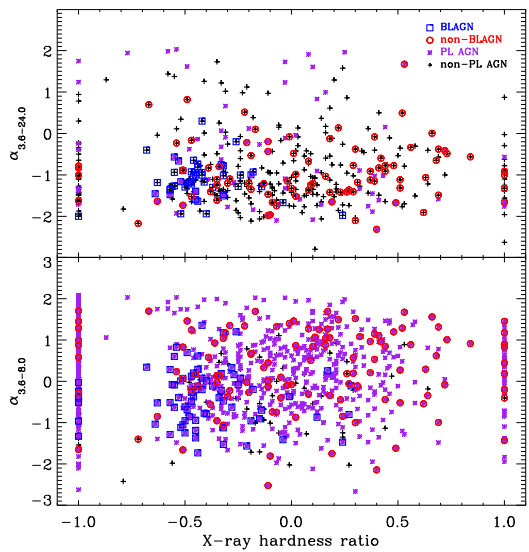}
\end{center}
\caption{Power-law indices $\alpha_{3.6-24 \mu m}$ and $\alpha_{3.6-8\mu m}$ (top and bottom panels, respectively) versus X-ray hardness ratio, $(H-S)/(H+S)$, where H represents the counts in the hard {\it Chandra} band (2-8 keV) and S represents the counts in the soft band (0.5-2 keV). 
Sources that are well fit by a power law over the given wavelength range ({\it purple $*$}) are indicated, other sources ({\it black $+$}) are plotted with a best-fit value of spectral slopes.
The sequence of points at -1.0 (+1.0) are from sources detected only in the soft (hard) band.  
There is no correlation between observed slopes at X-ray and infrared wavelengths.
The X-ray emission is most likely produced close to the black hole, while the mid-infrared emission likely comes from dust around the nucleus and in the host galaxy.}
\end{figure}

Our results are consistent with the findings for other X-ray-selected samples, where a significant fraction of the X-ray detected AGN have bluer colors indistinguishable from those of galaxies (e.g., \citealt{franceschinietal2005,barmbyetal2006}).
\citet{treisteretal06} concluded that a fair fraction of the mid-infrared light in low-luminosity AGN arises from the host galaxies, based on the large overlap in the infrared luminosity distributions of normal galaxies and lower luminosity AGN in the GOODS sample.
\citet{hickoxetal2007} showed that sources were moved into mid-infrared selection boxes by combining an AGN template with increasing strength to a galaxy template.
Mid-infrared selection techniques can miss a significant fraction of AGN if the high column densities significantly dim the central power-law emission and/or star light significantly dilutes the AGN continuum.

Mid-infrared selection techniques can uncover sources whose SED is dominated by the central AGN emission.
AGN undetected in {\it Chandra} or XMM surveys because of heavy obscuration are very likely to be bright infrared sources.
\citet{Alonso-Herrero2006} used the 24 $\mu$m-detected sample of sources in GOODS-South to select IRAC power-law galaxies.
Only half of their sample were individually detected at X-ray wavelengths.
\citet{pollettaetal2006} effectively used infrared-color selection techniques to uncover a sample of heavily obscured quasars.  
Their selection was based on featureless spectral energy distribution, and they deliberately removed any source that could be fit acceptably by a galaxy spectral template.
A complementary study by \citet{daddietal2007} examined non-X-ray-detected galaxies, which were not dominated by power-law continuum emission in the IRAC bands, and uncovered a detectable contribution from hot dust near an AGN in 20-30\% of the sample. 
This emission was discovered in excess mid-infrared flux above and beyond the amount predicted from star formation indicators at other wavelengths.
These galaxies lack any detectable signatures of AGN emission at optical wavelengths, but stacking results at X-ray wavelengths pointed towards obscured AGN emission \citep{daddietal2007}.

A well-defined subset of the X-ray counterparts detected at all infrared wavelengths are well-fit by red (negative) power laws from $3.6$ to 24  $\mu$m.
These sources have similar red power laws ($\alpha \le 0$), lie inside the standard infrared selection boxes for optically selected AGN, and show preferentially broad lines in their optical spectra.  
About 25\% of the whole sample fits into this category.
This is an upper bound since the shallow 24 $\mu$m coverage in the SWIRE survey may bias us towards redder X-ray sources.
To check this, we selected a sample of sources in the deeper GOODS area that are bright enough at 3.6 $\mu$m to be detected at 24 $\mu$m, even if they have a blue slope ($\alpha=+1$).
Of these 26 X-ray sources with bright 3.6 $\mu$m magnitudes ($m_{3.6} \le 21.4$), 4 have a  monotomic red slope between 3.6 and 24 $\mu$m.  This is a slightly lower fraction (15\%) but consistent within the statistics with the 25\% seen in the large sample.
These mid-infared power-law sources, tend to be the most luminous X-ray sources.
Together with their Compton-thick counterparts undetected at X-ray wavelengths, these are the best candidates for infrared selection techniques.

\section{\uppercase{Conclusions \& Summary}}
AGN are luminous infrared sources and deep infrared data from {\it Spitzer} potentially provide powerful new methods for selecting AGN samples.
In this paper, we examined the infrared properties of a large sample of AGN at moderately high redshifts selected through their X-ray emission.
In the observed 3.6 - 8.0 $\mu$m band over half of these X-ray-selected AGN show infrared colors consistent with those of galaxies.
Across the narrow wavelength range of the IRAC filters, most galaxies are also consistent with power-law spectral shapes.
As a result, IRAC data alone are not sufficient for a complete or reliable selection of (moderate luminosity) AGN.
At 24 $\mu$m, dust from both star forming galaxies and AGN is detected.
Unfortunately, observed IRAC to 24 $\mu$m colors do not effectively separate galaxies powered by AGN from those experiencing star formation.
Previous studies with IRAS have shown the far-infrared to be valuable for discriminating active from inactive galaxies \citep{sandersmirabel1996};
to apply similar techniques to present deep surveys requires longer wavelength data, not just at 24 $\mu$m but at 70 $\mu$m and if possible 160 $\mu$m, as well.

To develop complete samples of AGN, we need to understand what kinds of AGN each selection criterion returns.  
Infrared selection returns a slightly different population than X-ray selection.
Future studies using complete SEDs from the ultraviolet through infrared will help disentangle the contribution from the central AGN from that of the host galaxy.  
We have imaged the entire ECDF-S with medium-band filters, which will result in high quality photometric redshifts for most of the AGN and will allow us to disentangle the nuclear and host galaxy light.
These data will also reveal the origin of the variety of infrared colors shown below 8 $\mu$m.
The large range in color shown by AGN below 8 $\mu$m potentially holds valuable information about the star formation processes going on in the AGN host galaxies and may help us understand the AGN-galaxy connection.

\acknowledgements  We thank the anonymous referee for constructive comments that improved this paper.  Support from NSF grant \#AST0407295 and Spitzer JPL Grant \#RSA1288440 is gratefully acknowledged

\clearpage

\begin{landscape}
\tabletypesize{\scriptsize}
\begin{deluxetable}{rrrrrrrrrrrrr}
\tablecolumns{13}
\tablewidth{0pc}
\tablecaption{X-ray and Infrared Properties of MUSYC ECDF-S X-ray Sources \label{cat}}
\tabletypesize{\scriptsize}
\tablehead{
\colhead{ID} & \colhead{Cat ID$^{\rm a}$} & \colhead{RA} &  \colhead{DEC} & \colhead{HR$^{\rm b}$} & \colhead{$3.6 \mu m^{\rm c}$} & \colhead{$4.5 \mu m^{\rm c}$} & \colhead{$5.8 \mu m^{\rm c}$} & \colhead{$8.0 \mu m^{\rm c}$} & \colhead{$24.0 \mu m^{\rm d}$} & \colhead{${L_X}^{\rm e}$} & \colhead{Notes$^{\rm f}$}\\
\colhead{} & \colhead{} & \multicolumn{2}{c}{(J2000)} & \colhead{}  & \multicolumn{4}{c}{($\mu$Jy)} & \colhead{($\mu$Jy)} & \colhead{ (${\rm ergs \: s^{-1}}$)} & \colhead{} \\
\cline{3-4}  \cline{6-9} \\ 
}
\small
\startdata
1 & A03:1 &52.934208 &-27.823917 & -0.09 &   1.52 (0.14)  &    1.89 (0.15)  &   1.62 (0.62)  &  2.31 (0.57)  & \nodata (\nodata)  & \nodata &  \\
2 & A03:2 &52.936333 &-27.865639 & -0.43 &   6.69 (0.12)  &    7.94 (0.14)  &   5.33 (0.56)  &  5.98 (0.57)  & \nodata (\nodata)  & \nodata &  \\
3 & A03:3 &52.936875 &-27.860889 & -0.24 &  11.95 (0.13)  &   14.31 (0.15)  &  17.28 (0.56)  & 14.77 (0.58)  & \nodata (\nodata)  & \nodata &  \\
4 & A03:4 &52.947083 &-27.887028 & -0.34 &  15.15 (0.13)  &   22.73 (0.14)  &  32.26 (0.58)  & 46.54 (0.54)  & \nodata (\nodata)  & \nodata &  \\
5 & A03:5 &52.949917 &-27.845972 & -0.54 &   5.48 (0.14)  &    4.25 (0.16)  &   3.93 (0.61)  &  2.71 (0.62)  & \nodata (\nodata)  &  43.53 & 2 BL \\
6 & A03:6 &52.950083 &-27.800528 & -0.47 &  23.38 (0.13)  &   28.93 (0.15)  &  30.18 (0.62)  & 22.41 (0.60)  & S: 560.34 (28.32)  &  41.37 & 5  \\
7 & A03:7 &52.955958 &-27.776222 & -0.37 &  13.32 (0.13)  &   16.04 (0.16)  &  20.63 (0.59)  & 25.57 (0.63)  & \nodata (\nodata)  &  44.17 & 5  \\
8 & A03:8 &52.956208 &-27.842778 & -0.28 &   5.33 (0.14)  &    7.03 (0.16)  &   7.80 (0.61)  &  7.44 (0.62)  & \nodata (\nodata)  & \nodata &  \\
9 & A03:9 &52.959958 &-27.845028 &  0.10 &  15.81 (0.14)  &   19.24 (0.15)  &  15.07 (0.61)  & 10.65 (0.62)  & \nodata (\nodata)  & \nodata &  \\
10 & A03:10 &52.960125 &-27.864389 & -0.42 &12.34 (0.12)  &   11.83 (0.15)  &   8.86 (0.57)  &  7.60 (0.58)  & \nodata (\nodata)  & \nodata &  \\
11 & A03:11 &52.961000 &-27.883583 & -0.33 & 2.42 (0.13)  &    2.85 (0.15)  &   2.54 (0.59)  &  6.24 (0.59)  & \nodata (\nodata)  & \nodata &  \\
12 & A03:12 &52.963167 &-27.847667 & -0.29 &25.68 (0.14)  &   16.94 (0.16)  &  13.30 (0.60)  &  8.44 (0.62)  & \nodata (\nodata)  &  42.57 & 2 non-BL \\

\enddata

\tablecomments{This table is published in its entirety in the electronic edition of the Journal. A portion is shown here for guidance regarding its form and content.}
\tablenotetext{a}{Original detection catalog for X-ray source: and X-ray source ID there-in.  Counts and Flux are taken from this catalog.}
\tablenotetext{b}{Hardness ratio, defined by $\frac{H-S}{H+S}$, where H and S are the X-ray counts in the hard (2-8 keV) and soft (0.5-2 keV) bands, respectively.  For sources detected only in the hard band (soft band), the hardness ratio is set equal to 1 (-1).  For sources detected only in the full band (0.5-8 keV), the hardness ratio is set equal to 10.  G02 defined the hard band as 2-10 keV.  Therefore, we adjusted the published hard band counts from G02 to equal those that would have been detected at 2-8 keV, the hard band defined by the other three authors, using PIMMS (version 3.9c; \citealt{mukai1993}) with the average X-ray spectrum measured by G02.}
\tablenotetext{c}{Infrared flux in $\mu$Jy followed by 1$\sigma$ error in parentheses, from SIMPLE Survey (Damen et al., in prep)}
\tablenotetext{d}{24 micron flux in $\mu$Jy followed by 1$\sigma$ error in parentheses. Source of 24 micron flux is indicated by S for SWIRE \citep{lonsdale2003} or G for GOODS \citep{dickinson2003}}
\tablenotetext{e}{Observed luminosity [ergs/s] in the 0.5-8 keV band, using references given in final column.}
\tablenotetext{f}{Source of redshift used to calculate the luminosity: 0 for VLT/FORS2 Spectroscopy Version 1.0, 1 for VIMOS VLT Deep Survey (VVDS) Version 1.0, 2 for spectroscopic redshifts from \citet{szokolyetal2004}, 3 for spectroscopic redshifts from \citet{croometal2001}, 4 for spectroscopic redshifts determined from MUSYC spectroscopy (Treister et al. 2008, in prep.), 5 for photometric redshifts from COMBO-17 \citep{wolfetal2004}.  Where known, spectra showing broad emission lines (BL) are distinguished from those without broad lines (non-BL).}
\end{deluxetable}

\clearpage
\end{landscape}


\begin{thebibliography}{64}
\expandafter\ifx\csname natexlab\endcsname\relax\def\natexlab#1{#1}\fi

\bibitem[{{Alexander} {et~al.}(2003)}]{alexanderetal2003}
{Alexander}, D.~M. {et~al.} 2003, \aj, 126, 539

\bibitem[{{Alonso-Herrero} {et~al.}(2006)}]{Alonso-Herrero2006}
{Alonso-Herrero}, A. {et~al.} 2006, \apj, 640, 167

\bibitem[{{Barmby} {et~al.}(2006)}]{barmbyetal2006}
{Barmby}, P. {et~al.} 2006, \apj, 642, 126

\bibitem[{{Bertin} \& {Arnouts}(1996)}]{bertinarnouts1996}
{Bertin}, E. \& {Arnouts}, S. 1996, \aaps, 117, 393

\bibitem[{{Brandt} \& {Hasinger}(2005)}]{branthasinger2005}
{Brandt}, W.~N. \& {Hasinger}, G. 2005, \araa, 43, 827

\bibitem[{{Brinkmann} {et~al.}(2000){Brinkmann}, {Laurent-Muehleisen}, {Voges},
  {Siebert}, {Becker}, {Brotherton}, {White}, \& {Gregg}}]{brinkmannetal2000}
{Brinkmann}, W., {Laurent-Muehleisen}, S.~A., {Voges}, W., {Siebert}, J.,
  {Becker}, R.~H., {Brotherton}, M.~S., {White}, R.~L., \& {Gregg}, M.~D. 2000,
  \aap, 356, 445

\bibitem[{{Brusa} {et~al.}(2005)}]{brusa2005}
{Brusa}, M. {et~al.} 2005, \aap, 432, 69

\bibitem[{{Brusa} {et~al.}(2007)}]{brusaetal2007}
---. 2007, \apjs, 172, 353

\bibitem[{{Cardamone} {et~al.}(2007){Cardamone}, {Moran}, \&
  {Kay}}]{cardamoneetal2007}
{Cardamone}, C.~N., {Moran}, E.~C., \& {Kay}, L.~E. 2007, \aj, 134, 1263

\bibitem[{{Ciliegi} {et~al.}(2003){Ciliegi}, {Vignali}, {Comastri}, {Fiore},
  {La Franca}, \& {Perola}}]{ciliegietal2003}
{Ciliegi}, P., {Vignali}, C., {Comastri}, A., {Fiore}, F., {La Franca}, F., \&
  {Perola}, G.~C. 2003, \mnras, 342, 575

\bibitem[{{Croom} {et~al.}(2001){Croom}, {Warren}, \&
  {Glazebrook}}]{croometal2001}
{Croom}, S.~M., {Warren}, S.~J., \& {Glazebrook}, K. 2001, \mnras, 328, 150

\bibitem[{{Croton} {et~al.}(2006)}]{crotonetal2006}
{Croton}, D.~J. {et~al.} 2006, \mnras, 365, 11

\bibitem[{{Daddi} {et~al.}(2007)}]{daddietal2007}
{Daddi}, E. {et~al.} 2007, ArXiv e-prints, 705

\bibitem[{Damen}  {et~al.}(2007)]{damenetal2007}
{Damen}, M. {et~al.} 2008, in preparation

\bibitem[{{Dickinson} {et~al.}(2003){Dickinson}, {Giavalisco}, \& {The Goods
  Team}}]{dickinson2003}
{Dickinson}, M., {Giavalisco}, M., \& {The Goods Team}. 2003, in The Mass of
  Galaxies at Low and High Redshift, ed. R.~{Bender} \& A.~{Renzini}, 324--+

\bibitem[{{Donley} {et~al.}(2007){Donley}, {Rieke}, {P{\'e}rez-Gonz{\'a}lez},
  {Rigby}, \& {Alonso-Herrero}}]{donleyetal2007}
{Donley}, J.~L., {Rieke}, G.~H., {P{\'e}rez-Gonz{\'a}lez}, P.~G., {Rigby},
  J.~R., \& {Alonso-Herrero}, A. 2007, \apj, 660, 167

\bibitem[{{Dwelly} \& {Page}(2006)}]{dwellypage2006}
{Dwelly}, T. \& {Page}, M.~J. 2006, \mnras, 372, 1755

\bibitem[{{Egami} {et~al.}(2004)}]{egamietal2004}
{Egami}, E. {et~al.} 2004, \apjs, 154, 130

\bibitem[{{Elvis} {et~al.}(1994)}]{elvisetal1994}
{Elvis}, M. {et~al.} 1994, \apjs, 95, 1

\bibitem[{{Fabian} {et~al.}(2003){Fabian}, {Sanders}, {Allen}, {Crawford},
  {Iwasawa}, {Johnstone}, {Schmidt}, \& {Taylor}}]{fabian2003}
{Fabian}, A.~C., {Sanders}, J.~S., {Allen}, S.~W., {Crawford}, C.~S.,
  {Iwasawa}, K., {Johnstone}, R.~M., {Schmidt}, R.~W., \& {Taylor}, G.~B. 2003,
  \mnras, 344, L43

\bibitem[{{Fabian} {et~al.}(2006){Fabian}, {Sanders}, {Taylor}, {Allen},
  {Crawford}, {Johnstone}, \& {Iwasawa}}]{fabianetal2006}
{Fabian}, A.~C., {Sanders}, J.~S., {Taylor}, G.~B., {Allen}, S.~W., {Crawford},
  C.~S., {Johnstone}, R.~M., \& {Iwasawa}, K. 2006, \mnras, 366, 417

\bibitem[{{Fazio} {et~al.} (2004)}]{fazioetal2004}
{Fazio}, G., et al. 2004, \apjs, 154, 10

\bibitem[{{Franceschini} {et~al.}(2005)}]{franceschinietal2005}
{Franceschini}, A. {et~al.} 2005, \aj, 129, 2074

\bibitem[{{Freeman} {et~al.}(2002){Freeman}, {Kashyap}, {Rosner}, \&
  {Lamb}}]{freemanetal2002}
{Freeman}, P.~E., {Kashyap}, V., {Rosner}, R., \& {Lamb}, D.~Q. 2002, \apjs,
  138, 185

\bibitem[{{Gawiser} {et~al.}(2006)}]{gawiseretal2006}
{Gawiser}, E. {et~al.} 2006, \apjs, 162, 1

\bibitem[{{Georgantopoulos} {et~al.}(2007){Georgantopoulos}, {Georgakakis}, \&
  {Akylas}}]{georgantopoulosetal2007}
{Georgantopoulos}, I., {Georgakakis}, A., \& {Akylas}, A. 2007, \aap, 466, 823

\bibitem[{{Giacconi} {et~al.}(2002)}]{giacconietal2002}
{Giacconi}, R. {et~al.} 2002, \apjs, 139, 369

\bibitem[{{Granato} \& {Danese}(1994)}]{granatodanese1994}
{Granato}, G.~L. \& {Danese}, L. 1994, \mnras, 268, 235

\bibitem[{{Hatziminaoglou} {et~al.}(2005)}]{hatziminaolgou2005}
{Hatziminaoglou}, E. {et~al.} 2005, \aj, 129, 1198

\bibitem[{{Hickox} {et~al.}(2007)}]{hickoxetal2007}
{Hickox}, R.~C. {et~al.} 2007, ArXiv e-prints, 708

\bibitem[{{Kriek} {et~al.}(2006)}]{krieketal2006}
{Kriek}, M. {et~al.} 2006, ArXiv Astrophysics e-prints

\bibitem[{{Lacy} {et~al.}(2004)}]{lacyetal2004}
{Lacy}, M. {et~al.} 2004, \apjs, 154, 166

\bibitem[{{Laurent} {et~al.}(2000){Laurent}, {Mirabel}, {Charmandaris},
  {Gallais}, {Madden}, {Sauvage}, {Vigroux}, \& {Cesarsky}}]{laurentetal2000}
{Laurent}, O., {Mirabel}, I.~F., {Charmandaris}, V., {Gallais}, P., {Madden},
  S.~C., {Sauvage}, M., {Vigroux}, L., \& {Cesarsky}, C. 2000, \aap, 359, 887

\bibitem[{{Le F{\`e}vre} {et~al.}(2004)}]{lefevreetal2004}
{Le F{\`e}vre}, O. {et~al.} 2004, \aap, 428, 1043

\bibitem[{{Le Floc'h} {et~al.}(2004)}]{leflochetal2004}
{Le Floc'h}, E. {et~al.} 2004, \apjs, 154, 170

\bibitem[{{Lehmer} {et~al.}(2005)}]{lehmeretal2005}
{Lehmer}, B.~D. {et~al.} 2005, \apjs, 161, 21

\bibitem[{{Lira} {et~al.}(2002){Lira}, {Ward}, {Zezas}, {Alonso-Herrero}, \&
  {Ueno}}]{liraetal2002}
{Lira}, P., {Ward}, M., {Zezas}, A., {Alonso-Herrero}, A., \& {Ueno}, S. 2002,
  \mnras, 330, 259

\bibitem[{{Lonsdale} {et~al.}(2003)}]{lonsdale2003}
{Lonsdale}, C.~J. {et~al.} 2003, \pasp, 115, 897

\bibitem[{{Mazzarella} \& {Balzano}(1986)}]{Mazzarellabalzano1986}
{Mazzarella}, J.~M. \& {Balzano}, V.~A. 1986, \apjs, 62, 751

\bibitem[{{Moran} {et~al.}(2002){Moran}, {Filippenko}, \&
  {Chornock}}]{moranetal2002}
{Moran}, E.~C., {Filippenko}, A.~V., \& {Chornock}, R. 2002, \apjl, 579, L71

\bibitem[{{Mukai}(1993)}]{mukai1993}
{Mukai}, K. 1993, Legacy, vol.~3, p.21-31, 3, 21

\bibitem[{{Nenkova} {et~al.}(2002){Nenkova}, {Ivezi{\'c}}, \&
  {Elitzur}}]{nenkovaetal2002}
{Nenkova}, M., {Ivezi{\'c}}, {\v Z}., \& {Elitzur}, M. 2002, \apjl, 570, L9

\bibitem[{{Persic} {et~al.}(2004)}]{persic2004}
{Persic}, M. {et~al.} 2004, \aap, 427, 35

\bibitem[{{Pier} \& {Krolik}(1992)}]{pierkrolik1992}
{Pier}, E.~A. \& {Krolik}, J.~H. 1992, \apj, 401, 99

\bibitem[{{Polletta} {et~al.}(2006)}]{pollettaetal2006}
{Polletta}, M.~d.~C. {et~al.} 2006, \apj, 642, 673

\bibitem[{{Quadri} {et~al.}(2007)}]{quadrietal2007}
{Quadri}, R. {et~al.} 2007, \aj, 134, 1103

\bibitem[{{Richards} {et~al.}(2004)}]{richardsetal2004}
{Richards}, G.~T. {et~al.} 2004, \apjs, 155, 257

\bibitem[{{Rieke} {et~al}(2004)}]{riekeetal2004}
{Rieke}, G., {et al.} 2004, \apjs, 154, 25

\bibitem[{{Sanders}(1999)}]{sanders1999}
{Sanders}, D.~B. 1999, in IAU Symposium, Vol. 194, Activity in Galaxies and
  Related Phenomena, ed. Y.~{Terzian}, E.~{Khachikian}, \& D.~{Weedman}, 25--+

\bibitem[{{Sanders} \& {Mirabel}(1996)}]{sandersmirabel1996}
{Sanders}, D.~B. \& {Mirabel}, I.~F. 1996, \araa, 34, 749

\bibitem[{{Sanders} {et~al.}(1989){Sanders}, {Phinney}, {Neugebauer}, {Soifer},
  \& {Matthews}}]{sandersetal1989}
{Sanders}, D.~B., {Phinney}, E.~S., {Neugebauer}, G., {Soifer}, B.~T., \&
  {Matthews}, K. 1989, \apj, 347, 29

\bibitem[{{Schawinski} {et~al.}(2007){Schawinski}, {Thomas}, {Sarzi},
  {Maraston}, {Kaviraj}, {Joo}, {Yi}, \& {Silk}}]{schawinskietal2007}
{Schawinski}, K., {Thomas}, D., {Sarzi}, M., {Maraston}, C., {Kaviraj}, S.,
  {Joo}, S.-J., {Yi}, S.~K., \& {Silk}, J. 2007, ArXiv e-prints, 709

\bibitem[{{Severgnini} {et~al.}(2003)}]{severgninietal2003}
{Severgnini}, P. {et~al.} 2003, \aap, 406, 483

\bibitem[{{Sijacki} {et~al.}(2007){Sijacki}, {Springel}, {di Matteo}, \&
  {Hernquist}}]{sijackietal2007}
{Sijacki}, D., {Springel}, V., {di Matteo}, T., \& {Hernquist}, L. 2007,
  \mnras, 380, 877

\bibitem[{{Stern} {et~al.}(2005)}]{sternetal2005}
{Stern}, D. {et~al.} 2005, \apj, 631, 163

\bibitem[{{Streblyanska} {et~al.}(2004){Streblyanska}, {Bergeron}, {Brunner},
  {Finoguenov}, {Hasinger}, \& {Mainieri}}]{streblyanska2004}
{Streblyanska}, A., {Bergeron}, J., {Brunner}, H., {Finoguenov}, A.,
  {Hasinger}, G., \& {Mainieri}, V. 2004, Nuclear Physics B Proceedings
  Supplements, 132, 232

\bibitem[{{Streblyanska} {et~al.}(2006){Streblyanska}, {Bergeron}, {Brunner},
  {Finoguenov}, {Hasinger}, \& {Mainieri}}]{streblyanska2006}
{Streblyanska}, A., {Bergeron}, J., {Brunner}, H., {Finoguenov}, A.,
  {Hasinger}, G., \& {Mainieri}, V. 2006, in IAU Symposium, Vol. 230,
  Populations of High Energy Sources in Galaxies, ed. E.~J.~A. {Meurs} \&
  G.~{Fabbiano}, 450--454

\bibitem[{{Surace} {et~al.}(2005){Surace}, {Shupe}, {Fang}, {Evans}, {Alexov},
  {Frayer}, {Lonsdale}, \& {SWIRE Team}}]{surace2005}
{Surace}, J.~A., {Shupe}, D.~L., {Fang}, F., {Evans}, T., {Alexov}, A.,
  {Frayer}, D., {Lonsdale}, C.~J., \& {SWIRE Team}. 2005, in Bulletin of the
  American Astronomical Society, Vol.~37, Bulletin of the American Astronomical
  Society, 1246--+

\bibitem[{{Sutherland} \& {Saunders}(1992)}]{sutherlandsaunders1992}
{Sutherland}, W. \& {Saunders}, W. 1992, \mnras, 259, 413

\bibitem[{{Szokoly} {et~al.}(2004)}]{szokolyetal2004}
{Szokoly}, G.~P. {et~al.} 2004, \apjs, 155, 271

\bibitem[{Treister}  {et~al.}(2008)]{treisteretal2008}
{Treister}, E. {et~al.} 2008, in preparation

\bibitem[{{Treister} \& {Urry}(2006)}]{treisterurry2006}
{Treister}, E. \& {Urry}, C.~M. 2006, \apjl, 652, L79

\bibitem[{{Treister} {et~al.}(2004)}]{treisteretal2004}
{Treister}, E. {et~al.} 2004, \apj, 616, 123

\bibitem[{{Treister} {et~al.}(2006)}]{treisteretal06}
{Treister}, E. {et~al.} 2006, \apj, 640, 603

\bibitem[{{Vanzella} {et~al.}(2005)}]{vanzella2005}
{Vanzella}, E. {et~al.} 2005, \aap, 434, 53

\bibitem[{{Virani} {et~al.}(2006){Virani}, {Treister}, {Urry}, \&
  {Gawiser}}]{viranietal2006}
{Virani}, S.~N., {Treister}, E., {Urry}, C.~M., \& {Gawiser}, E. 2006, \aj,
  131, 2373

\bibitem[{{Wolf} {et~al.}(2004)}]{wolfetal2004}
{Wolf}, C. {et~al.} 2004, \aap, 421, 913

\bibitem[{{Worsley} {et~al.}(2005)}]{worsleyetal2005}
{Worsley}, M.~A. {et~al.} 2005, \mnras, 357, 1281

\end{thebibliography}
\end{document}